\documentclass[fleqn,twoside]{article}
\usepackage{amsmath,amssymb,bm,curves}
\usepackage{espcrc2}
\usepackage{graphics, color, cite, amsmath, rotating, epic}
\usepackage{eepic}

\def\bge{\begin{equation}}
\def\ene{\end{equation}}
\def\bg{\begin{eqnarray}}
\def\en{\end{eqnarray}}

\def\ubar{{\bar{u}}}
\def\dbar{{\bar{d}}}
\def\sbar{{\bar{s}}}

\def\vr{\vec{r}}

\title{$\eta$ bound states in nuclei: a probe of flavour-singlet dynamics}

\author{Steven D.\ Bass\address{Institute for Theoretical Physics, 
Universit\"at Innsbruck,
Technikerstrasse 25, A 6020 Innsbruck, Austria}%
\ and
Anthony W.\ Thomas\address%
{Jefferson Laboratory, 12000 Jefferson Avenue,
Newport News, VA 23606, U.S.A.}%
}

\begin{document}

\begin{abstract}
We argue that
$\eta$ bound states in nuclei are sensitive to the singlet component
in the $\eta$. The bigger the singlet component, the more attraction
and the greater the binding.
Thus, measurements of $\eta$ bound states will yield new information
about axial U(1) dynamics and glue in mesons.
$\eta - \eta'$ mixing plays an important role 
in understanding the value of the $\eta$-nucleon scattering length.

\vspace{1pc}
\end{abstract}

\maketitle

\section{INTRODUCTION}

Measurements of the pion, kaon and eta meson masses and their interactions
in finite nuclei provide new constraints on our understanding of dynamical
symmetry breaking in low energy QCD \cite{kienle}.
New experiments at the GSI will employ the recoilless $(d, \ ^3He)$ 
reaction to study the possible formation of $\eta$ meson bound states 
inside the nucleus \cite{hayano},
following on from the successful studies of pionic atoms in these reactions
\cite{pionexpt}.
The idea is to measure the excitation-energy spectrum and then, if a clear 
bound state is observed,
to extract the in-medium effective mass, $m_{\eta}^*$, of the $\eta$
in nuclei through performing a fit to this spectrum with the $\eta$-nucleus
optical potential.

In this paper we argue that $m_{\eta}^*$ is sensitive to the flavour-singlet 
component in the $\eta$, and hence to non-perturbative glue 
\cite{uppsala,zuoz} associated with axial U(1) dynamics.
An important source of the in-medium mass modification comes 
from light-quarks
coupling to the scalar $\sigma$ mean-field in the nucleus.
Increasing the flavour-singlet component in the $\eta$
at the expense of the octet component gives more attraction, 
more binding and a larger value of the $\eta$-nucleon 
scattering length, $a_{\eta N}$.
This result may explain why values of $a_{\eta N}$ extracted from 
phenomenological fits to experimental data where the $\eta-\eta'$ 
mixing angle is unconstrained 
give larger values than those predicted 
in theoretical models where the $\eta$ is treated as a pure octet state.

We first introduce the basic physics. 
Next, in Section 2 we briefly review the QCD axial U(1) problem and its 
application to the $\eta$ mass in nuclei.
We motivate the {\it existence} of gluonic corrections to $m_{\eta}^*$
which go beyond pure Goldstone boson dynamics.
While QCD arguments imply information about the sign of the mass shift,
a rigorous numerical calculation of $m_{\eta}^*$ from QCD is presently 
not feasible.
Hence, in Section 3, we consider QCD inspired model predictions for the 
$\eta$ and $\eta'$-nucleus systems and the vital role of flavour-singlet 
degrees of freedom in $\eta$ bound-states.
In Section 4 we summarize and conclude.

Meson masses in nuclei are determined from the scalar induced contribution 
to the meson propagator evaluated at zero three-momentum, ${\vec k} =0$, in 
the nuclear medium.
Let $k=(E,{\vec k})$ and $m$ denote the four-momentum and mass of the meson 
in free space.
Then, one solves the equation
\begin{equation}
k^2 - m^2 = {\tt Re} \ \Pi (E, {\vec k}, \rho)
\end{equation}
for ${\vec k}=0$
where $\Pi$ is the in-medium $s$-wave meson self-energy.
Contributions to the in medium mass come from coupling to the scalar 
$\sigma$ field in the nucleus in mean-field approximation,
nucleon-hole and resonance-hole excitations in the medium.
The $s$-wave self-energy can be written as \cite{ericson}
\begin{equation}
\Pi (E, {\vec k}, \rho) \bigg|_{\{{\vec k}=0\}}
=
- 4 \pi \rho \biggl( { b \over 1 + b \langle {1 \over r} \rangle } \biggr) .
\end{equation}
Here $\rho$ is the nuclear density,
$
b = a ( 1 + {m \over M} )
$
where 
$a$ is the meson-nucleon scattering length, $M$ is the nucleon mass and
$\langle {1 \over r} \rangle$ is
the inverse correlation length,
$\langle {1 \over r} \rangle \simeq m_{\pi}$ 
for nuclear matter density \cite{ericson}.
($m_{\pi}$ is the pion mass.) 
Attraction corresponds to positive values of $a$.
The denominator in Eq.(2) is the Ericson-Ericson-Lorentz-Lorenz
 double scattering correction.

What should we expect for the $\eta$ and $\eta'$ ?

\section{QCD CONSIDERATIONS}

Spontaneous chiral symmetry breaking is associated with a non-vanishing
chiral condensate
\begin{equation}
\langle \ {\rm vac} \ | \ {\bar q} q \ | \ {\rm vac} \ \rangle < 0
.
\label{eq5}
\end{equation}
The non-vanishing chiral condensate also spontaneously breaks the axial
U(1) symmetry so, naively, in the two-flavour theory one expects an
isosinglet pseudoscalar degenerate with the pion.
The lightest mass isosinglet is the $\eta$ meson, which has a mass of
547.75 MeV.

The puzzle deepens when one considers SU(3).
Spontaneous chiral symmetry breaking suggests an octet of 
would-be Goldstone bosons:
the octet associated with chiral $SU(3)_L \otimes SU(3)_R$
plus a singlet boson associated with axial U(1)
--- each with mass squared $m^2_{\rm Goldstone} \sim m_q$.
The physical $\eta$ and $\eta'$ masses
are
about 300-400 MeV too big to fit in this picture.
One needs extra mass in the singlet channel
associated with
non-perturbative topological gluon configurations and
the QCD axial anomaly \cite{zuoz}.
The strange quark mass induces considerable $\eta$-$\eta'$ mixing.
For free mesons
the $\eta - \eta'$ mass matrix (at leading order in the chiral
expansion) is
\begin{equation}
M^2 =
\left(\begin{array}{cc}
{4 \over 3} m_{\rm K}^2 - {1 \over 3} m_{\pi}^2  &
- {2 \over 3} \sqrt{2} (m_{\rm K}^2 - m_{\pi}^2) \\
\\
- {2 \over 3} \sqrt{2} (m_{\rm K}^2 - m_{\pi}^2) &
[ {2 \over 3} m_{\rm K}^2 + {1 \over 3} m_{\pi}^2 + {\tilde m}^2_{\eta_0} ]
\end{array}\right)
.
\label{eq10}
\end{equation}
Here ${\tilde m}^2_{\eta_0}$ is the gluonic mass term which has a
rigorous interpretation through the Witten-Veneziano mass formula
\cite{wv,vecca}
and which
is associated with non-perturbative gluon
topology, related perhaps to confinement \cite{ks} or instantons
\cite{thooft}.
The masses of the physical $\eta$ and $\eta'$ mesons are found
by diagonalizing this matrix, {\it viz.}
\begin{eqnarray}
| \eta \rangle &=&
\cos \theta \ | \eta_8 \rangle - \sin \theta \ | \eta_0 \rangle
\\ \nonumber
| \eta' \rangle &=&
\sin \theta \ | \eta_8 \rangle + \cos \theta \ | \eta_0 \rangle
\label{eq11}
\end{eqnarray}
where
\begin{equation}
\eta_0 = \frac{1}{\sqrt{3}}\; (u\ubar + d\dbar + s\sbar),\quad
\eta_8 = \frac{1}{\sqrt{6}}\; (u\ubar + d\dbar - 2 s\sbar) 
.
\label{mixing2}
\end{equation}
One obtains values for the $\eta$ and $\eta'$ masses:
\begin{eqnarray}
m^2_{\eta', \eta} 
& &= (m_{\rm K}^2 + {\tilde m}_{\eta_0}^2 /2)
\nonumber \\
& & \pm {1 \over 2}
\sqrt{(2 m_{\rm K}^2 - 2 m_{\pi}^2 - {1 \over 3} {\tilde m}_{\eta_0}^2)^2
   + {8 \over 9} {\tilde m}_{\eta_0}^4} 
.
\nonumber \\
\label{eq12}
\end{eqnarray}
The physical mass of the $\eta$ and the octet mass
$
m_{\eta_8} = \sqrt{ {4 \over 3} m_{\rm K}^2 - {1 \over 3} m_{\pi}^2 }
$
are numerically close, within a few percent.
However, to build a theory of the $\eta$ on the octet
approximation
risks losing essential physics associated with the singlet component.
Turning off the gluonic term, one finds the expressions
$m_{\eta'} \sim \sqrt{2 m_{\rm K}^2 - m_{\pi}^2}$
and
$m_{\eta} \sim m_{\pi}$.
That is, without extra input from glue, in the OZI limit,
the $\eta$ would be approximately an isosinglet light-quark state
(${1 \over \sqrt{2}} | {\bar u} u + {\bar d} d \rangle$)
degenerate with the pion and
the $\eta'$ would be a strange-quark state $| {\bar s} s \rangle$
--- mirroring the isoscalar vector $\omega$ and $\phi$ mesons.

Taking the value ${\tilde m}_{\eta_0}^2 = 0.73$GeV$^2$ in the 
leading-order 
mass formula, Eq.(\ref{eq12}), 
gives agreement with the physical masses at the 10\% level.
This value is obtained by summing over the two eigenvalues 
in Eq.(7):
$
m_{\eta}^2 + m_{\eta'}^2 = 2 m_K^2 + {\tilde m}_{\eta_0}^2 
$
and substituting 
the physical values of $m_{\eta}$, $m_{\eta'}$ and $m_K$ \cite{vecca}.
The 
corresponding
$\eta - \eta'$
mixing angle $\theta \simeq - 18^\circ$ 
is within the range $-17^\circ$ to $-20^\circ$ obtained
from a study of various decay processes in \cite{gilman,frere}.
\footnote{
Closer agreement with the physical masses can be obtained 
by introducing
the singlet decay constant
$F_0 \neq F_{\pi}$ and including higher-order mass terms in the chiral
expansion
\cite{leutwyler,feldmann}.
}
The key point of Eq.(7) is that mixing and gluon dynamics play a crucial 
role 
in both the $\eta$ and $\eta'$ masses
and 
that treating the $\eta$ as an octet pure would-be Goldstone boson risks 
losing essential physics.

\subsection{$\eta$ and $\eta'$ interactions with the nuclear medium}

What can QCD tell us about the behaviour of the gluonic mass contribution 
in the nuclear medium ?

The physics of axial U(1) degrees of freedom is described 
by the 
U(1)-extended low-energy effective Lagrangian \cite{vecca}.
In its simplest form this reads
\begin{eqnarray}
{\cal L} =
{F_{\pi}^2 \over 4}
{\rm Tr}(\partial^{\mu}U \partial_{\mu}U^{\dagger})
+
{F_{\pi}^2 \over 4} {\rm Tr} M \biggl( U + U^{\dagger} \biggr)
\nonumber \\
+ {1 \over 2} i Q {\rm Tr} \biggl[ \log U - \log U^{\dagger} \biggr]
+ {3 \over {\tilde m}_{\eta_0}^2 F_{0}^2} Q^2
.
\nonumber \\
\label{eq20}
\end{eqnarray}
Here 
$
U = \exp \ i \biggl(  \phi / F_{\pi}
                  + \sqrt{2 \over 3} \eta_0 / F_0 \biggr) 
$
is the unitary meson matrix
where
$\phi = \ \sum \pi_a \lambda_a$ 
denotes the octet of would-be Goldstone bosons associated 
with spontaneous chiral $SU(3)_L \otimes SU(3)_R$ breaking
and
$\eta_0$
is the singlet boson.
In Eq.(8) $Q$ denotes the topological charge density
($Q = {\alpha_s \over 4 \pi} G_{\mu \nu} {\tilde G}^{\mu \nu}$);
$M = {\rm diag} [ m_{\pi}^2, m_{\pi}^2, 2 m_K^2 - m_{\pi}^2 ]$
is the quark-mass induced meson mass matrix.
The pion decay constant $F_{\pi} = 92.4$MeV and 
$F_0$ is
the flavour-singlet decay constant,
$F_0 \sim F_{\pi} \sim 100$ MeV \cite{gilman}.

The flavour-singlet potential involving $Q$ is introduced to generate 
the gluonic contribution to the $\eta$ and $\eta'$ masses and
to reproduce the anomaly in the divergence of
the gauge-invariantly renormalized flavour-singlet axial-vector
current.
The gluonic term $Q$ is treated as a background field with no kinetic 
term. It may be eliminated through its equation of motion to generate 
a gluonic mass term for the singlet boson,
{\it viz.}
\begin{equation}
{1 \over 2} i Q {\rm Tr} \biggl[ \log U - \log U^{\dagger} \biggr]
+ {3 \over {\tilde m}_{\eta_0}^2 F_{0}^2} Q^2
\
\mapsto \
- {1 \over 2} {\tilde m}_{\eta_0}^2 \eta_0^2
.
\label{eq23}
\end{equation}
The most general low-energy effective Lagrangian involves a $U_A(1)$
invariant polynomial in $Q^2$.  Higher-order terms in $Q^2$ become
important when we consider scattering processes involving more than
one $\eta'$ \cite{veccb}.
In general, couplings involving $Q$ give OZI violation in physical
observables.

To investigate what happens to ${\tilde m}^2_{\eta_0}$ in the medium
we first couple
the $\sigma$ 
(correlated two-pion)
mean-field in nuclei
to the topological charge density $Q$.
The interactions of the $\eta$ and $\eta'$ with other mesons and 
with nucleons can be studied by coupling the Lagrangian Eq.(8) to 
other particles.
For example,
the OZI violating interaction
$\lambda Q^2 \partial_{\mu} \pi_a \partial^{\mu} \pi_a$
is needed to generate the leading (tree-level)
contribution to the decay $\eta' \rightarrow \eta \pi \pi$
\cite{veccb}.
When iterated in the Bethe-Salpeter equation for meson-meson
rescattering
this interaction yields a dynamically generated exotic state
with quantum numbers $J^{PC} = 1^{-+}$ and mass about 1400 MeV
\cite{bassmarco}.
This suggests a dynamical interpretation of the lightest-mass 
$1^{-+}$ exotic observed at BNL and CERN.

Motivated by this two-pion coupling to $Q^2$, we couple the topological 
charge density to the $\sigma$ (two-pion) mean-field in the nucleus
by adding the Lagrangian term
\begin{equation}
{\cal L}_{\sigma Q} =
Q^2 \ g_{\sigma}^Q \sigma
\label{eq27}
\end{equation}
where 
$g_{\sigma}^Q$ denotes coupling to the $\sigma$ mean field
--
that is, we
consider an in-medium renormalization of the coefficient of $Q^2$
in the effective chiral Lagrangian.
Following the treatment in Eq.(9) we eliminate
$Q$ through its equation of motion.
The gluonic mass term for the singlet boson then becomes
\begin{equation}
{\tilde m}^2_{\eta_0}
\mapsto
{\tilde m}^{*2}_{\eta_0}
=
{\tilde m}^2_{\eta_0}
\ { 1 + 2 x \over (1 + x)^2 }
\ < {\tilde m}^2_{\eta_0}
\label{eq28}
\end{equation}
where
\begin{equation}
x =
{1 \over 3} g_{\sigma}^Q \sigma \ {\tilde m}^2_{\eta_0} F_0^2.
\label{eq29}
\end{equation}
That is, {\it the gluonic mass term decreases in-medium}
independent of the sign of $g_{\sigma}^Q$ and the medium acts
to partially neutralize axial U(1) symmetry breaking by gluonic effects.

This scenario has possible support from recent lattice calculations 
\cite{derek}
which suggest that non-trivial gluon topology configurations are 
suppressed inside hadrons.
Further recent work at high chemical potential
$(\mu > 500 {\rm MeV})$
suggests that possible confinement and instanton
contributions to ${\tilde m}_{\eta_0}^2$
are suppressed with increasing density in this domain \cite{schafer}.
We investigate the size of the $\eta$ mass shift in Section 3 below.

\subsection{The $\eta$ nucleon scattering length and anomalous glue}

Further insight is provided from looking at the scattering length.
When the U(1)-extended chiral Lagrangian is coupled to nucleons one
finds new OZI violating couplings in the flavour-singlet sector
\cite{sb99}.
An example is the gluonic contribution to the singlet
Goldberger-Treiman relation \cite{shorev} which 
connects axial U(1)
dynamics and the spin structure of the proton studied
in polarized deep inelastic scattering and high-energy polarized
proton-proton collisions -- for a recent review see \cite{spin}.
In the chiral limit the singlet 
analogy to the Weinberg-Tomozawa
term does not vanish because of the anomalous glue terms.
Starting from the simple Born term one finds 
anomalous gluonic contributions 
to the singlet-meson nucleon scattering length
proportional to ${\tilde m}^2_{\eta_0}$ and ${\tilde m}_{\eta_0}^4$
\cite{bassww}.

We briefly summarize this section.

The masses of the $\eta$ and $\eta'$ receive contributions from 
terms associated with both explicit chiral symmetry breaking and
with anomalous glue through the Witten-Veneziano term.
Mixing is important and, ideally, one would like
to consider the medium dependence of the different basic physics inputs.
At the QCD level, OZI-violating gluonic couplings have the potential
to affect the effective $\eta$ and $\eta'$ masses in nuclei and,
through Eq.(2), the $\eta$-nucleon and $\eta'$-nucleon scattering lengths.
It is interesting to also mention the observation of Brodsky et al. 
\cite{brodsky} that attractive gluonic van der Waals type exchanges 
have the potential to produce flavour-singlet $\eta_c$ bound-states 
in the $(d, \ ^3He)$ reaction close to threshold.

The above discussion is intended to motivate the {\it existence} of 
medium modifications to ${\tilde m}^2_{\eta_0}$ in QCD.
However, a rigorous calculation of $m_{\eta}^{*}$ from QCD 
is beyond present theoretical technology. 
Hence, one has to look to QCD motivated models and phenomenology for
guidance about the numerical size of the effect.
The physics described in 
Eqs.(4-7) tells us that the simple octet approximation may not suffice.

\section{MODELS}

We now discuss the size of flavour-singlet effects in $m_{\eta}^*$, 
$m^*_{\eta'}$ 
(the $\eta'$ mass in-medium) 
and the scattering lengths $a_{\eta N}$ and $a_{\eta' N}$.
First we consider the values of 
$a_{\eta N}$ and $a_{\eta' N}$ 
extracted from phenomenological fits to experimental data.
There are several model predictions for the $\eta$ mass in 
nuclear matter,
starting from different assumptions.
We collect and compare these approaches and predictions
with particular emphasis on the contribution of $\eta - \eta'$ mixing.
We also compare model predictions for the internal structure of the 
$S_{11}(1535)$ nucleon resonance and its in-medium excitation energy.

{\it Phenomenological determinations of $a_{\eta N}$ and $a_{\eta' N}$:}
Green and Wycech \cite{wycech} have performed phenomenological 
K-matrix 
fits to a variety of near-threshold processes 
($\pi N \rightarrow \pi N$, $\pi N \rightarrow \eta N$,
 $\gamma N \rightarrow \pi N$ and $\gamma N \rightarrow \eta N$)
to extract a value for the $\eta$-nucleon scattering.
%
In these fits the $S_{11}(1535)$ is introduced as an explicit
degree of freedom
-- that is, it is treated like a 3-quark state --
and the $\eta-\eta'$ mixing angle is taken as a free parameter.
The real part of $a_{\eta N}$ extracted from these fits is 0.91(6) fm
for the on-shell scattering amplitude.

From measurements of $\eta$ production in proton-proton collisions 
close to threshold, 
COSY-11 have extracted a scattering length 
$a_{\eta N} \simeq 0.7 + i 0.4$fm 
from the final state interaction (FSI)
based on the effective range approximation
 \cite{cosyeta}.
For the $\eta'$, COSY-11 have deduced a
conservative upper bound on 
the $\eta'$-nucleon scattering length
$| {\tt Re} a_{\eta' N} | < 0.8$fm \cite{cosy}
with a prefered a value between 0 and 0.1 fm \cite{pawel}
obtained by comparing the FSI in $\pi^0$ and $\eta'$ production 
in proton-proton collisions close to threshold.

{\it Chiral Models:}
Chiral models involve performing a coupled channels analysis of 
$\eta$ production after multiple rescattering in the nucleus
which is calculated
using the Lippmann-Schwinger \cite{etaweise} or Bethe-Salpeter 
\cite{etaoset} equations with potentials taken from the SU(3)
chiral Lagrangian for low-energy QCD.
In these chiral model calculations
the $\eta$ is taken as pure octet state
$(\eta = \eta_8)$ with no mixing and the singlet sector turned off.
These calculations
yield a small mass shift in nuclear matter
\begin{equation}
m^*_{\eta} / m_{\eta} \simeq 1 - 0.05 \rho / \rho_0
\end{equation}
The values of the $\eta$-nucleon scattering length extracted from
these chiral model calculations are
0.2 + i 0.26 fm
\cite{etaweise} and
0.26 + i 0.24 fm
\cite{etaoset}
with slightly different treatment of the intermediate state mesons.

{\it The Quark Meson Coupling Model:}
The third approach we consider is the Quark-Meson Coupling model (QMC)
\cite{etaqmc}.
Here one uses the large $\eta$ mass 
(which in QCD is induced by mixing and the gluonic mass term)
to motivate taking an MIT Bag 
description
for the $\eta$ wavefunction, and
then coupling the light (up and down)
quark and antiquark fields in the $\eta$ to the scalar $\sigma$
field
in the nucleus working in mean-field approximation \cite{etaqmc}.
The strange-quark component of the wavefunction does not couple
to the $\sigma$ field 
and
$\eta-\eta'$ mixing is readily built into the model.

The mass for the $\eta$ in nuclear matter is self-consistently 
calculated by solving for the MIT Bag in the nuclear medium \cite{etaqmc}:
\begin{equation}
m_\eta^*(\vr) = \frac{2 [a_P^2\Omega_q^*(\vr)
+ b_P^2\Omega_s(\vr)] - z_\eta}{R_\eta^*}
+ {4\over 3}\pi R_\eta^{* 3} B,
\label{meta}\\
\end{equation}
\begin{equation}
\left.\frac{\partial m_j^*(\vr)}
{\partial R_j}\right|_{R_j = R_j^*} = 0, \quad\quad (j = \eta, \ \eta') .
\label{equil}\\
\end{equation}
Here $\Omega_q^*$ and $\Omega_s$ are light-quark and strange-quark
Bag energy eigenvalues,
$R_{\eta}^*$ is the Bag radius in the medium and $B$ is the Bag constant.
The $\eta-\eta'$ mixing angle $\theta$ is included in the terms
$
a_P = \frac{1}{\sqrt{3}} \cos\theta
- \sqrt{\frac{2}{3}} \sin\theta
$
and
$
b_P = \sqrt{\frac{2}{3}} \cos\theta
+ \frac{1}{\sqrt{3}} \sin\theta
$.
and can be varied in the model.
One first solves the Bag for the free $\eta$ with a 
given mixing angle, and then turns on QMC to obtain the mass-shift.
Results for the $\eta'$ are obtained by interchanging 
$a_P \leftrightarrow b_P$.
In Eq.~(\ref{meta}), $z_\eta$ parameterizes the sum of the center-of-mass 
and gluon fluctuation effects, and is assumed to be independent of
density~\cite{finite0}.
The current quark masses are taken as $m_q = 5$ MeV and $m_s = 250$ MeV.
\footnote{
This is the strange-quark mass needed to reproduce the Lambda and Sigma 
masses 
in the model. 
While larger than the values for $m_s$ quoted at momentum scales relevant 
perturbative QCD, 
the Bag model approximates QCD at a very low scale 
(well below 1 GeV)
- a region where renormalization group
evolution would make the running masses much larger than at 2 GeV$^2$. }

The coupling constants in the model for the coupling of light-quarks 
to the $\sigma$ (and $\omega$ and $\rho$) mean-fields in the nucleus 
are 
adjusted to fit the saturation energy and density of 
symmetric nuclear matter and the bulk symmetry energy.
The Bag parameters used in these calculations are
$\Omega_q = 2.05$
(for the light quarks)
and $\Omega_s = 2.5$
(for the strange quark) 
for free hadrons
with
$B = ( 170 {\rm MeV} )^4$.
For nuclear matter density we find $\Omega^*_q = 1.81$ for the 1$s$ state.
This value depends on the coupling of light-quarks to the $\sigma$ 
mean-field and is independent of the mixing angle $\theta$.
Likewise, $\Omega_q$ and $\Omega_s$ are determined by solving for
light and strange quarks in the MIT Bag potential and are independent of
$\theta$.

For the $\eta$ and $\eta'$ mesons the $\omega$ vector mean-field couples 
with the same magnitude and opposite sign to the quarks and antiquarks 
in the meson, and therefore cancels. 
Increasing the mixing angle increases the amount of singlet 
relative to octet components in the $\eta$.
This produces greater attraction through increasing 
the amount of light-quark compared to strange-quark 
components in the $\eta$
and a reduced effective mass.
Through Eq.(2) increasing the mixing angle 
also increases the 
$\eta$-nucleon scattering length $a_{\eta N}$.
We quantify this in Table 1 which presents results 
for the pure octet 
($\eta=\eta_8$, $\theta=0$) 
and the values 
$\theta = - 10^\circ$ and $- 20^\circ$ (the physical mixing angle).

\begin{table}[htbp]
\begin{center}
\caption{
Physical masses fitted in free space, 
the bag 
masses in medium at normal nuclear-matter 
density,
$\rho_0 = 0.15$ fm$^{-3}$, 
and corresponding meson-nucleon scattering lengths (see below).
}
\label{bagparam}
\begin{tabular}[t]{c|lll}
\hline
&$m$ (MeV) 
& $m^*$ (MeV) & ${\tt Re} a$ (fm)
\\
\hline
$\eta_8$  &547.75  
& 500.0 &  0.43 \\
$\eta$ (-10$^o$)& 547.75  
& 474.7 & 0.64 \\
$\eta$ (-20$^o$)& 547.75  
& 449.3 & 0.85 \\
$\eta_0$  &      958 
& 878.6  & 0.99 \\
$\eta'$ (-10$^o$)&958 
& 899.2 & 0.74 \\
$\eta'$ (-20$^o$)&958 
& 921.3 & 0.47 \\
\hline
\end{tabular}
\end{center}
\end{table}
The values of ${\tt Re} a_{\eta}$ quoted in Table 1 are obtained
from substituting the in-medium and free masses into Eq.(2) with
the Ericson-Ericson denominator turned-off, and using the free
mass $m=m_{\eta}$ 
in the expression for $b$. The effect of exchanging $m$ for
$m^*$ in $b$ is a 5\% increase in the quoted scattering length.
The QMC model makes no claim about the imaginary part of the scattering 
length. 
The key observation is that $\eta - \eta'$ mixing leads to a factor of 
two increase in the mass-shift and in the scattering length obtained in 
the model.

The QMC model is calibrated by fixing the coupling constants to the observed 
properties of nuclear matter or finite nuclei.
So, even though it is mean-field (no correlations) it does fit observed 
binding energies. When one applies the same model with the same (quark level 
couplings) to the binding of etas the natural belief is that it should 
give the physical binding energies.
From these one can extract an effective scattering length.
Because the QMC model has been explored mainly at the mean-field level, 
it is not clear that one should include the Ericson-Ericson-Lorentz-Lorenz
term in extracting the corresponding $\eta$ nucleon scattering length. 
If one substitutes the scattering lengths given in Table 1 
into Eq.(2) 
(and neglects the imaginary part which is not predicted by the model)
one obtains resummed values 
$a_{eff} = a / ( 1 + b  \langle 1/r \rangle )$ 
equal to 0.44 fm for the $\eta$ and 0.28 fm 
for the $\eta'$
for the physical mixing angle $\theta = -20$ degrees.
(Here we take $\langle 1/r \rangle \simeq m_{\pi}$ for nuclear matter
 density \cite{ericson}.)

The density dependence of the mass-shifts in the QMC model is discussed
in Ref.\cite{etaqmc}.
Neglecting the Ericson-Ericson term, the mass-shift is approximately
linear. For densities $\rho$ between 0.5 and 1 times $\rho_0$ (nuclear
matter density) we find
\begin{equation}
m^*_{\eta} / m_{\eta} \simeq 1 - 0.17 \rho / \rho_0
\end{equation}
for the physical mixing angle $-20^\circ$.
The scattering lengths extracted from this analysis are density independent 
to within a few percent over the same range of densities.

Finally, we note that in the QMC treatment 
one assumes that the value of the mixing angle does not 
change in medium. As
mentioned above this is not excluded and merits 
further investigation.

{\it The $S_{11}(1535)$ resonance in nuclear matter:}
It is interesting to compare the different model predictions for the 
$S_{11}(1535)$ nucleon resonance which couples strongly to the 
$\eta$-nucleon system.
\footnote{
We refer to \cite{jido} for a recent discussion of the role of 
the $S_{11}(1535)$ in the $\eta$-nucleus optical potential.}
In quark models the $S_{11}$ is interpreted as a 3-quark state: $(1s)^2(1p)$.
This interpretation has support from quenched lattice calculations
\cite{lattice}
which also suggest that the $\Lambda (1405)$ resonance has a significant non 
3-quark component. 
In the Cloudy Bag Model the $\Lambda (1405)$ is dynamically generated in the 
kaon-nucleon system \cite{CBM}.
Chiral coupled channels models with an octet $\eta = \eta_8$ agree with 
these predictions for the $\Lambda (1405)$ and differ for the $S_{11} (1535)$,
which is interpreted as a $K \Sigma$ quasi-bound state \cite{kaiser}.

Experiments in heavy-ion collisions 
\cite{averbeck} 
and $\eta$ photoproduction
from nuclei \cite{robig,yorita} suggest little modification of 
the $S_{11} (1535)$ excitation in-medium, though some evidence for the 
broadening of the $S_{11}$ in nuclei was reported in \cite{yorita}.
Despite the different physics input, both QMC and the coupled channels 
models agree with this finding.
In QMC the excitation energy is $\sim 1544$ MeV.
This is obtained as follows.
For a quark in the 1p state the Bag light-quark energy-eigenvalue in free 
space is $\Omega_q = 3.81$.
In QMC at normal nuclear-matter density this is reduced to $\Omega^*_q = 3.77$.
(Note the smaller mass shift compared to the s-wave eigenvalue.)
The scalar mass term for the $S_{11}$ is reduced to $\sim 1424$ MeV through 
coupling to the $\sigma$ mean-field.
The scalar attraction is compensated by repulsion from coupling 
to the omega mean-field, $\sim +120$ MeV, to give the excitation 
energy 1544 MeV.
(For the $\eta$,
 the $\omega$ mean-field coupling to the quark and antiquark
 enters
 with equal magnitude and opposite sign and therefore cancels.)
In chiral coupled channels calculations one finds a similar 
$S_{11}$ excitation energy $ \sim 1560$ MeV.
Here the medium independence of the resonance excitation energy is 
interpreted as arising from the absence of Pauli blocking of the 
$K \Sigma$ system in nuclear matter.
We note that in QMC for all baryons the scalar attraction very nearly 
cancels the vector repulsion, leaving a small (few 10's of MeV) net 
attraction or repulsion.

\section{CONCLUSIONS}

$\eta$-$\eta'$ mixing increases the flavour-singlet and light-quark
components in the $\eta$.
The greater the flavour-singlet component in the $\eta$, 
the greater the $\eta$ binding energy in nuclei through 
increased attraction and the smaller the value of $m_{\eta}^*$.
Through Eq.(2), this corresponds to an increased $\eta$-nucleon 
scattering length $a_{\eta N}$,
greater than the value one would expect if the $\eta$ were a pure octet state.
Measurements of $\eta$ bound-states in nuclei 
are therefore a probe of singlet axial U(1) 
dynamics in the $\eta$.

It will be very interesting to see the results from the new GSI experiment
for $\eta$ bound-states.
Additional studies might be possible using $\eta$ 
production in low-energy proton-nucleus collisions and 
in photoproduction.
Here one might use an electromagnetic calorimeter, 
for example WASA@COSY,
to tag the two-photon decay of the $\eta$.
However, unlike the GSI programme, one has to be careful 
in these experiments 
whether the $\eta$ is produced inside the nucleus or on the surface.
Possibilities to study 
$\eta$ and 
$\eta'$-mesic nuclei in $(\gamma, p)$ spectra are discussed in \cite{spring}.

\section*{Acknowledgments}

We thank P. Kienle, D. Leinweber, P. Moskal and S. Wycech for helpful
communications.
The work of SDB is supported by the Austrian Research Fund, FWF,
through contract P17778.
The work of AWT is supported by DOE contract DE-AC05-84ER40159,
under which SURA operates Jefferson National Laboratory.


\end{document}